\documentclass[%
 reprint,
 amsmath,amssymb,
 aps,
pre,longbibliography]{revtex4-2}

\usepackage{blkarray}
\usepackage{amsmath}
\usepackage{graphicx}
\usepackage{dcolumn}
\usepackage{bm}
\usepackage{hyperref}

\usepackage{xcolor}

\begin{document}


\title{Dynamics--structure interchangeability in binary opinion models on multiplex networks}

 \author{Julian Sienkiewicz$^{1,2}$ and Anna Chmiel$^2$}
 \affiliation{$^1$Centre for Credible AI, Warsaw University of Technology, Rektorska 4, 00-614 Warszawa, Poland}
 \affiliation{$^2$Faculty of Physics, Warsaw University of Technology, Koszykowa 75, 00-662 Warszawa, Poland}






\begin{abstract}
We examine the interchangeability between two duplex systems with opinion dynamics: one characterized by different dynamics parameters $a_1$ and $a_2$ on each layer and full node overlap, and the second with the same parameter $a_1$ on both layers but only partial node overlap $r$. We introduce a general framework for identifying the critical line between the ordered and disordered phases, thereby translating one system into another. The description is supported by several examples of opinion dynamics models and a rigorous measure of the distance between the stationary solutions of the rate equations in both settings. Our results show that even large values of $r$ are sufficient to obtain very good correspondence between the two settings.    
\end{abstract}

\maketitle

\section{INTRODUCTION}
Multilevel and multiplex networks \cite{Boccaletti2014, DeDomenico2013} clearly belong to the complex systems area -- their introduction was directly connected to the fact that describing relations among the entities of a system with a single, aggregated network (a monoplex) can be insufficient to understand its topology \cite{Kurant2006}. The level of multilayer network structural reducibility, i.e., how many layers can be aggregated from the topological point of view, is not trivial, though, and relies heavily on the importance of specific layers \cite{DeDomenico2015}. This issue becomes even more important when we consider processes on the network, such as opinion dynamics \cite{Starnini2026,Diakonova2016}.

In this paper, we ask a different question: to what extent is dynamical heterogeneity between two fully overlapping layers — differing only in their parameters — equivalent to structural randomization of the overlap itself? To achieve this goal, we investigate a two-layer full-graph multiplex (duplex) with an opinion dynamics characterized by different parameters in each layer and juxtapose it with a duplex characterized by the same parameters in both layers, but subject to partial structural overlap. We condition these two systems on the critical line between the ordered and disordered states and test the theoretical description with specific examples of opinion dynamics.

Although we focus on complex systems examples, and more specifically on opinion models, the idea of layered coupled graphs is not limited to this area. A vivid example is the experimental study of two disordered superconductors \cite{Bonamassa2023} -- in this system, cross-layer coupling leads to a discontinuous phase transition in the resistance, rather than a continuous one, when the layers are isolated. Similarly, the overlap between layers, introduced as a layer rotation, substantially changes transport parameters in van der Waals thin-film systems \cite{Kim2021}. From this point of view, exploring the possibilities of imitating dynamical heterogeneity with a structural overlap could be extremely important.

The rest of this paper is organized as follows: we first introduce the general framework, and then describe the examined cases of opinion dynamics. We show how introducing heterogeneity changes these systems as opposed to monoplex and homogeneous settings. Then, based on the critical line conditions, we express the parameters of dynamics in one system via the structural overlap in the second one. We quantify the differences based on a rigorous metric and examine their change as a function of the overlap, discussing the obtained results.
\begin{figure}[!ht]
    \includegraphics[width=\columnwidth]{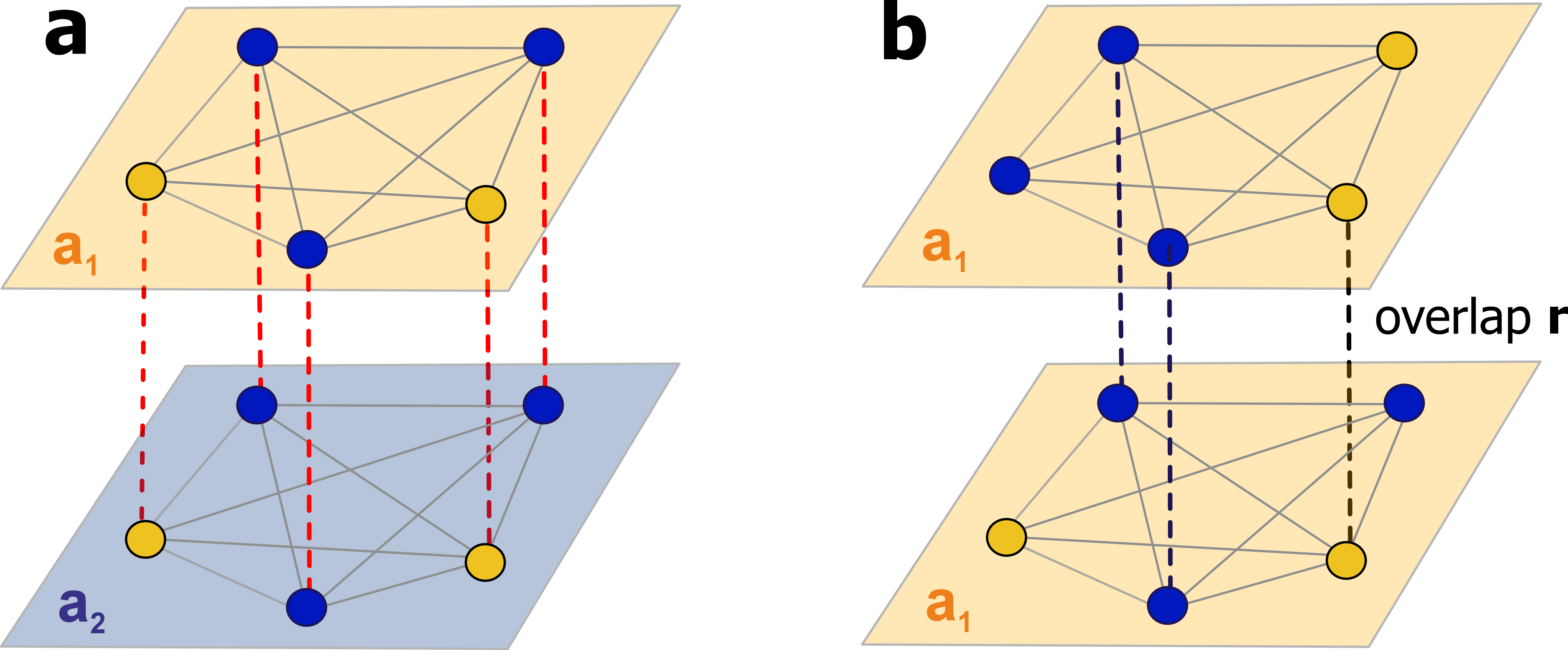}
    \caption{Schematic representation of examined settings: (a) a fully overlapped system governed with different dynamical parameters $a_1$ and $a_2$, (b) system with partial overlap $r$ governed by the same parameter $a_1$ in both layers.}
\label{fig:fig1}
\end{figure}

\section{Model-agnostic theoretical background} 
The generic system consists of $N$ agents, each holding a binary state $s_i=\pm1$, and it can be described by total magnetization $m=\frac{1}{N}\sum_{i=0}^{i=N}s_i$. We assume that the process that takes place on a fully connected graph and is characterized by some parameter $a$ and that the transition probability to change the node state from $-1$ to $+1$ is $\gamma^+(m,a) \equiv \gamma^+(a)$ while the opposite situation occurs with symmetrical probability $\gamma^-(a) \equiv \gamma^-(m,a) \equiv \gamma^+(-m,a)$. 

\begin{table*}[!t]
    \centering
    \begin{tabular}{cccc}
    \hline\hline
      dynamics  & transition probability $\gamma$ & role of the $a$ parameter & additional parameter\\
      \hline
      q-voter \cite{Nyczka2012} & $\gamma_V=(1-p)\left(\frac{1+m}{2}\right)^q+\frac{p}{2}$ & $p \in [0,1]$ - independence probability & $q > 0$\footnote{Typically it is tacitly assumed that the $q \in \mathbb{N}$ due to sociological aspects, however the theory works for any $q \in \mathbb{R}^+$.} - lobby size\\
      q-Ising\footnote{$E_{qk}=\min\{1,\exp[2(q-2k)/T]\}$} \cite{Jedrzejewski2015} & $\gamma_I=\sum_{k=0}^q{q \choose k}(\frac{1+m}{2})^{q-k}(\frac{1-m}{2})^k E_{qk}$& $T > 0$ - pseudotemperature & $q > 0$ - lobby size\\
      q-MV\footnote{Transition probability for odd values of $q$. The $I_z(a,b)$ is a regularized beta function \cite{Abramowitz1964}.} \cite{Nyczka2018} & $\gamma_{MV}=f+(1-2f)I_{(1+m)/2}\left(\frac{q+1}{2},\frac{q+1}{2}\right)$ & $f \in [0,1]$ - bias & $q > 0$ - lobby size\\
      Sznajd \cite{Sznajd-Weron2011} & $\gamma_{S}=(1-p)\left(\frac{1+m}{2}\right)^2+p f$ & $p \in [0,1]$ - independence probability & $f \in [0,1]$ - flexibility of opinion\\
      \hline\hline
    \end{tabular}
            \caption{Main properties of binary opinion models used in this study.}
    \label{tab:tab1}
\end{table*}

Now we introduce two extensions of this system in the framework of a two-layer multiplex network. The first one, denoted further as $G\equiv G(a_1,a_2)$ and depicted in Fig.~\ref{fig:fig1}a, consists of two full graphs connected in such a way that each node from one layer has its counterpart in the second one. However, each layer is characterized by a different value of the $a$ parameter: $a_1$ in the first one and $a_2$ in the second, therefore we will call it a heterogeneous system. The coupling of the layers is implemented via a so-called \texttt{AND} rule \cite{Lee2014}: the node changes its state only if the change is indicated on both levels. Given the above, the rate equation $F \equiv F(m,a_1,a_2)=0$ governing the coupled system for $N\rightarrow\infty$ reads
\begin{equation}
    F = \frac{1-m}{2}\gamma^+(a_1)\gamma^+(a_2)-\frac{1+m}{2}\gamma^-(a_1)\gamma^-(a_2) = 0.    
\label{eq:rate}
\end{equation}

If $G$ exhibits a phase transition between the ordered and disordered state, we can examine its properties by follothe wing Landau approach, i.e., by calculating the effective potential $V=-\int F dm$ and expanding it into a power series, keeping only the first three terms
\begin{equation}
    V(m) = Am^2 + Bm^4 + Cm^6.
\label{eq:landau}
\end{equation}
In particular, by solving $A=0$ (which is equivalent to $(\partial_m F)_{m=0}=0$) we arrive at the critical line equation $a_2 = f(a_1)$ if an explicit solution is possible.

Let us now describe the system $G_r \equiv G_r(a_1,r)$ presented in Fig.~\ref{fig:fig1}b: in contrast to the previous one, in each step of the dynamics, only a fraction $r$  of all nodes have their counterpart in the second layer -- they are called duplex nodes. The remaining $1-r$ nodes are monoplex ones, i.e., they do not follow the \texttt{AND} dynamics, but update their state taking into account only their layer. Both layers are subject to dynamics with the same parameter $a_1$ (identical to $a_1$ in $G$). Thus in this case we can write the rate equation $F_r \equiv F(m,a_1,r)=0$ as 
\begin{equation}
\begin{split}
F_r &= (1-r)\left[\frac{1-m}{2}\gamma^+(a_1)-\frac{1+m}{2}\gamma^-(a_1)\right] +\\
& r\left[\frac{1-m}{2}(\gamma^+(a_1))^2-\frac{1+m}{2}(\gamma^-(a_1))^2\right]=0.
\end{split}
\label{eq:dm}
\end{equation}

The proposed probabilistic dynamics is a case of an {\it annealed} system, unlike the {\it quenched} case proposed in \cite{Chmiel2017}. The latter relies on a specific structure that does not change, while the former assumes that there is probability $r$ to draw a duplex node and $1-r$ for the monoplex case, $F_r$ being a weighted sum of relevant effective forces $F_r=(1-r)F_{\rm monoplex}(m,a_1)+rF_{\rm duplex}(m,a_1,a_1)$. In a similar way as in $G$, we can find the critical line $a_1 = g(r)$ by solving $A=0$. Given both critical line relations, we are now able to formally translate parameter $a_2$ in the system $G_1$ as $r$ coming from $G_2$ with
\begin{equation}
\begin{aligned}
a_2(r) = f(g(r))
\end{aligned}
\label{eq:a2r}
\end{equation}
which is the key equation in this study, serving as a bridge between $G$ and $G_r$.

\section{Examples and properties of heterogeneous systems}
Given the general theory, we will now present and discuss the behavior of $G$ for specific examples, using four previously established models of opinion dynamics: the q-voter model with independence \cite{Nyczka2012} (later called q-voter), the q-neighbor Ising (later called q-Ising) model \cite{Jedrzejewski2015}, and the Sznajd model with independence \cite{Sznajd-Weron2011} (later called simply Sznajd for brevity) and a modified version of majority vote dynamics (q-MV model, which is a special case of the threshold q-voter model \cite{Nyczka2018}). Although some of these models have been investigated on the topology of multiplex full graphs or networks \cite{Chmiel2015, Chmiel2017, Gradowski2020, Krawiecki2023}, the idea of heterogeneous layers (in the sense of dynamical parameters) has not yet been examined, the closest being two processes (q-voter and q-Ising) in different layers \cite{Chmiel2025}. The key component of all the models is the \textit{q-lobby} (\textit{q-panel}) -- a set of $q$ nodes selected at random from all neighbors of the given agent $i$ (here, given the complete graph, out of all nodes). The idea of the lobby is grounded in a series of experiments performed by Solomon Asch \cite{Asch1955,Asch1956} showing the impact of a unanimous group on an individual. Transition probabilities $\gamma(m,a)$, together with a description of the parameters of the examined models, are presented in Table \ref{tab:tab1}. 
\begin{figure*}[]
         \includegraphics[width=.85\textwidth]{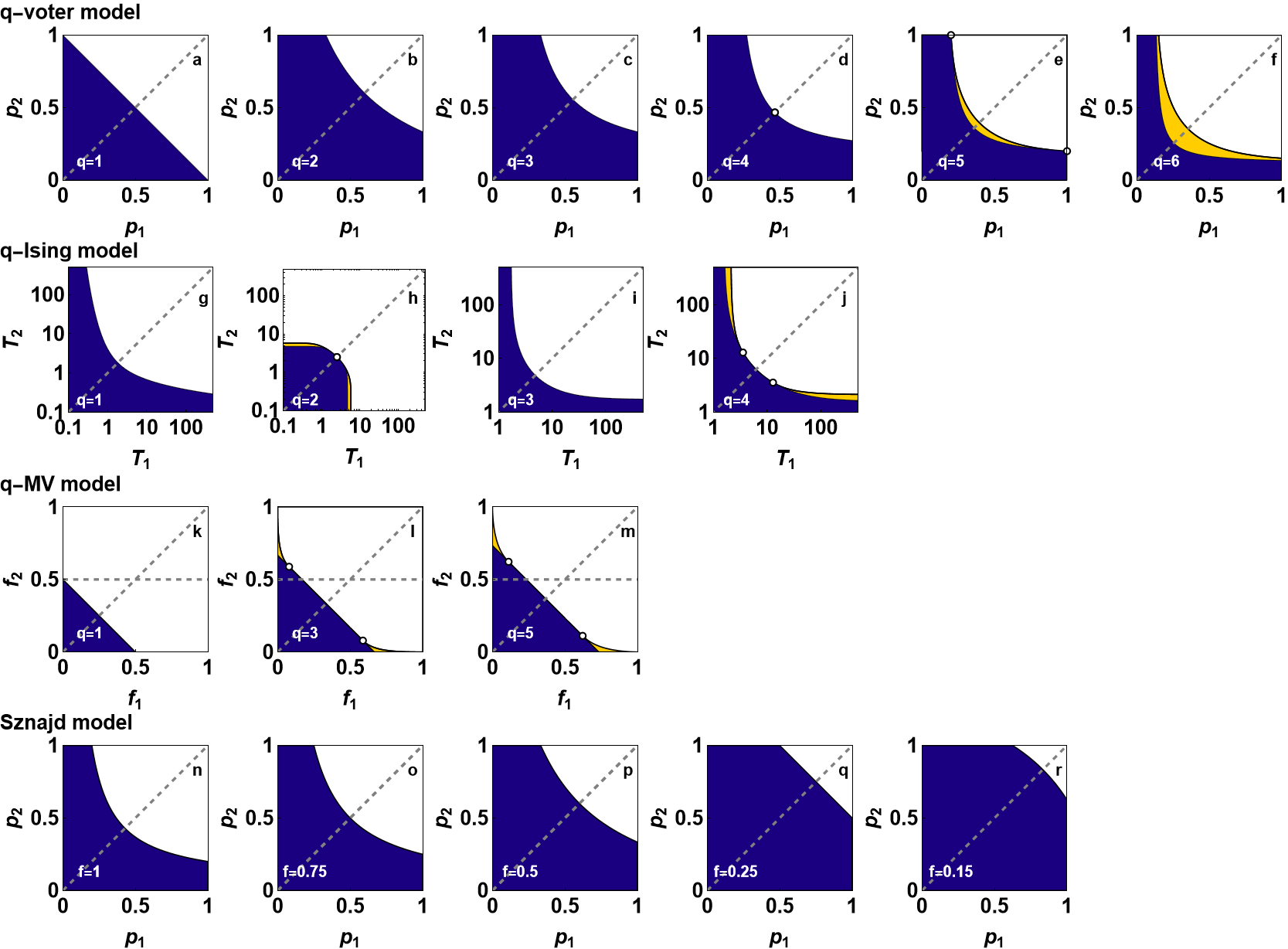}
\caption{Phase diagrams of heterogeneous q-voter model (a-f), q-Ising model (g-j), q-MV model (k-m), and the Sznajd model (n-r) for selected values of parameters $q$ and $f$. Dark regions indicate an ordered phase, white regions -- an unordered one. Orange regions mark the area of bistability of these two phases. Circles mark tricritical points. The diagonal dashed line represents phase diagram for a homogeneous duplex, while the horizontal line in the case of the q-MV model --- phase diagram for a monoplex. }
\label{fig:fig2}
\end{figure*}

In the case of the q-voter dynamics \cite{Nyczka2012}, the agent flips its state only when the $q$-lobby is homogeneous; however, it still has the option to change its opinion on its own, which takes place with probability $p$, called the probability of independence. Previous studies have shown that such a system is disordered for $q=1$, undergoes a continuous phase transition for $1 < q \le 5$ (tricritical at $q=5$) and a discontinuous one with a hysteresis for $q>5$ \cite{Nyczka2012}. The introduction of the \texttt{AND} coupling shifts this picture: a homogeneous duplex is characterized by a continuous phase transition for $1 \leq q \leq 4$ (tricritical at $q=4$) and a discontinuous one when $q>4$ \cite{Chmiel2015}. Results for the heterogeneous duplex setting $G(p_1,p_2)$ are shown in Fig. \ref{fig:fig2}a-f and confirm dependence on the panel size $q$, however, they can be deduced from the analysis of the monoplex and homogeneous duplex systems.

The situation is different for the q-Ising model. In analogy to the original Ising model, the opinion of the agent is flipped with probability $\min\{1,\exp(\Delta E/T)\}$, where $T>0$ is temperaturelike parameter, $\Delta E = 2s_i\sum_{nn}s_j$ and summation goes over all nodes in the q-lobby. The monoplex model is characterized by an unordered state for $q=1$ and $q=2$, a continuous phase transition for $q=3$, and a discontinuous one if $q \ge 4$ \cite{Jedrzejewski2015}. In the homogeneous duplex setting, reported numerical results indicate continuous phase transitions in the range of $q\in[1,6]$ \cite{Chmiel2017}. As seen in Fig. \ref{fig:fig2}g-j the introduction of the heterogeneity brings surprising results: the phase diagram for $q=1$ is very unlike the one for $q=2$, the former system undergoing a continuous phase transition for any temperature other than $T_1\rightarrow\infty$ and later confined in the region $T_1 \in (0;4/\ln2)$. Phase diagrams for $q=3$ and $q=4$, as in the q-voter case, can be deduced from the monoplex and homogeneous duplex settings.
\begin{figure*}[!ht]
         \includegraphics[width=.85\textwidth]{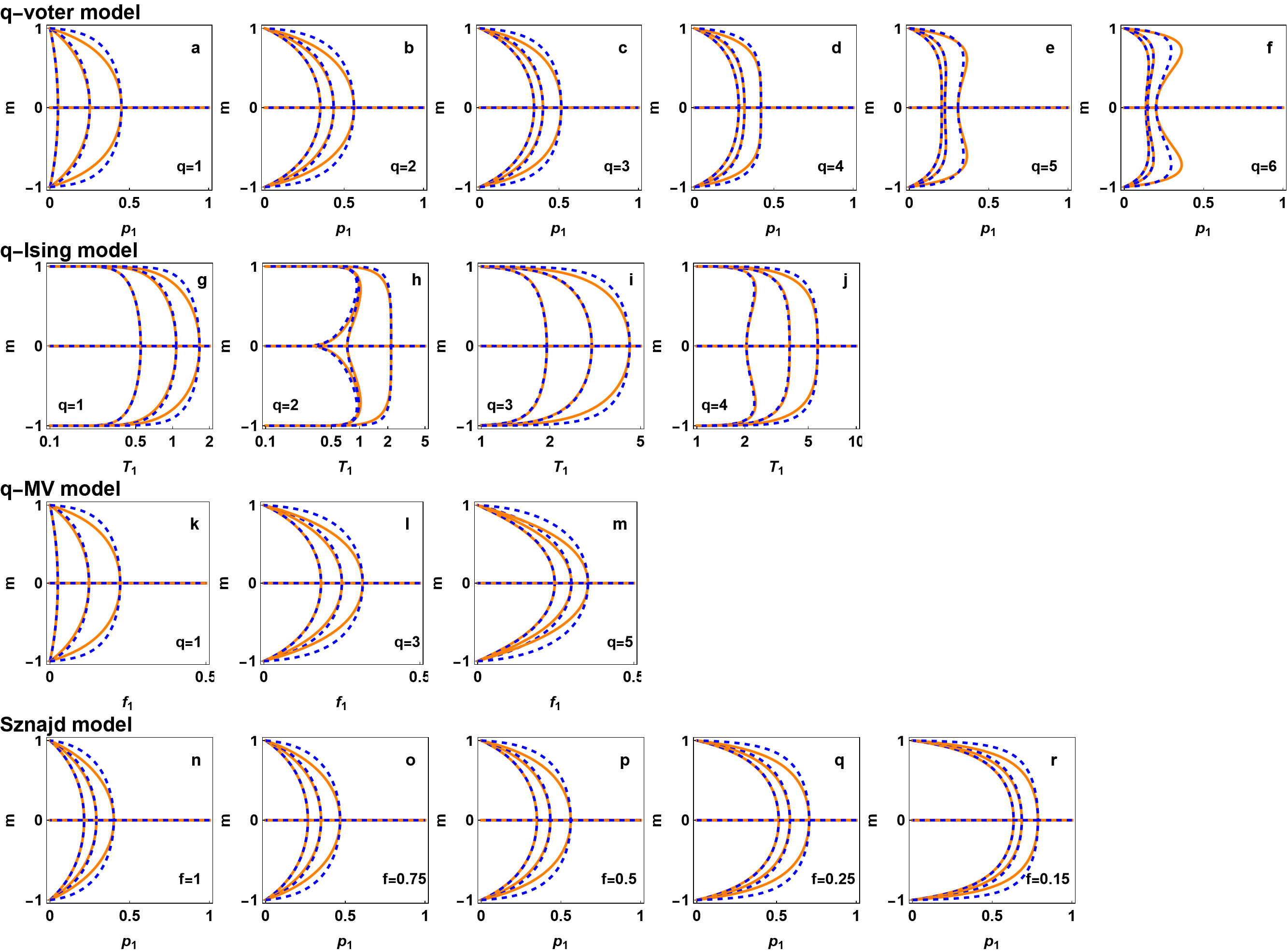}
\caption{Solutions of $F_r(m,a_1,r)=0$ with $F(m,a_1,a_2(r))=0$ for parameters $q$ (q-voter, q-Ising and q-MV) and $f$ (Sznajd) used previously in Fig. \ref{fig:fig2} and for specific values of $r$ ($r=\{4/7,0.58,0.9\}$ in the case of q-Ising model for $q=2$ and $r=\{0.1,0.5,0.9\}$ in any other system)}
\label{fig:fig3}
\end{figure*}

Heterogeneity of the dynamics parameters plays a crucial role in the q-MV model. The original majority vote model \cite{deOliveira1992} assumes that at each time step a chosen node adopts the majority opinion of the neighborhood with probability $1-f$ and the minority with $f$. Taking into account limited use for a full graph, we use a special case of the threshold q-voter model \cite{Nyczka2018} with $z=1$ and $r=\lceil (q+1)/2 \rceil$ and now the change of opinion is possible with $1-f$ according to the majority in the q-panel and with $f$ according to the minority in the same lobby. This leads to the following flip probability $\gamma_{MV}=2^{-q}\sum_{k=0}^q{q \choose k}\left(1+m\right)^{q-k}\left(1-m\right)^k E_{qk}$ where $E_{qk} = 1-f$ if $k < q/2$, $E_{qk} = f$ if $k > q/2$ and $E_{qk} = 1/2$ in the case of tie (for even values of $q$). For odd values of $q$, we can get a compound form in terms of the regularized beta function (see Table \ref{tab:tab1}). Following detailed calculations based on (\ref{eq:landau}), one can show that both monoplex and homogeneous duplex systems undergo continuous phase transitions except for a fully disordered monoplex $q=1$ (which is equivalent to the q-voter model $q=1$ with a rescaled parameter $p=2f$). The heterogeneous coupling (see Fig. \ref{fig:fig2}k-m) leads to an unexpected emergence of bistability for $q=3$, which is then maintained for larger panel sizes.    

The last examined system is the Sznajd model with independence \cite{Sznajd-Weron2011}, where in each time step two random nodes $i$ and $j$ influence the third node $k$: with probability $1-p$ the node changes is a conformist and changes its state to $s_i$ if $s_i=s_j$, and with $p$ it acts independently, changing its sign to $-s_k$ with probability $f$. Given the rules, we can consider the model to be equipped with the lobby of size 2 (in fact, $f=1/2$ version is equivalent to the $q=2$ q-voter model). The monoplex version undergoes a continuous phase transition with $p_c=1/(1+4f)$ \cite{Sznajd-Weron2011}, while the homogeneous duplex is characterized by $p_c=3/(3+4f)$ (also only continuous transition).  Phase diagrams for the heterogeneous version (Fig. \ref{fig:fig2}n-r) do not differ from the monoplex and duplex cases in terms of the character of the phase transition. 

\section{Interchangeability results}
Let us now move to the key results of this study. Given the forms of the transition probabilities and the overall theoretical background, we can obtain the relation (\ref{eq:a2r}) for the systems under examination, it is possible to derive closed forms for the three models
\begin{equation}
p^{V}_2(r)=\frac{4 \left(2^q (q-1) (r-1)-2 q (r-2)+r-2\right)}{R-(8 q-4) (r-2)+4^q (r-1)+2^{q+1} (q-1) (2 r-3)}
\end{equation}
where $R=\sqrt{4^q \left(4 q^2+\left(2^q-2\right) \left(4 q+2^q-2\right) (r-1)^2\right)}$,
\begin{equation}
f^{MV}_2(r)=\frac{1}{2}-r\frac{2^{q+1}\left(\frac{q+1}{2}\right)! \left(\frac{q-1}{2}\right)!}{8 (q+1)!}
\end{equation}
and 
\begin{equation}
\begin{aligned}
p^{S}_2(r) = & \frac{3 f+4 r-3-7 f r}{r (f+6-4 (f-2) r-9)+3 (f-1)-2 r R}
\end{aligned}
\end{equation}
with $R = \sqrt{16 (2f^2-2f+1) r^2+(f-1)^2-8 \left(f^2-1\right) r}$. Due to nonlinear involvement of $T$ in the $\gamma$ function for the q-Ising model it is impossible to provide a general formula in this system, however the two simplest cases read $T^{q=1}_2(r)=2/{\ln \left[(\sqrt{r^2-r+1}+r+1)/(\sqrt{r^2-r+1}-r+1)\right]}$ and $T^{q=2}_2(r)=4/\ln \left[(2 \sqrt{9 r^2-2 r+9}+6 r+1)/(7-4 r)\right]$.

The boundary values of the obtained relations are in line with the construction of $G$ and $G_r$. For $r=0$, $G_r$ consists of two separated layers, each characterized by parameter $a_1$. The realization of such a setting in the case of $G$ is only possible when one of the layers does not influence the other one, i.e., when the dynamics is purely random. This is recovered by $p_2=1$ for the q-voter and Sznajd models, $f_2=1/2$ for the q-MV, and $T_2\rightarrow\infty$ for the q-Ising, which can also be obtained by setting $r=0$ in the above relations. The second limiting situation ($r=1$) means that $G_r$ is a fully heterogeneous duplex. Based on our construction involving the critical line, the value of $p_2(r=1)$ is the critical value of $p_1$ in a fully heterogeneous duplex.

\begin{figure}[!bt]
         \includegraphics[width=\columnwidth]{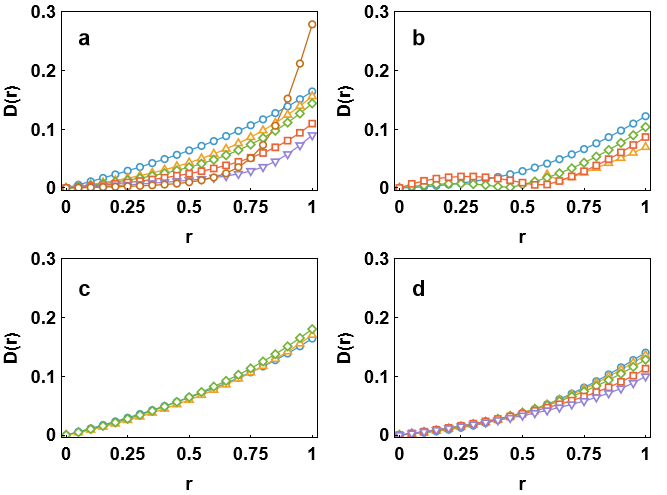}
\caption{Relative distance $D(r)$ against overlap $r$. Each point is a numerical integration of Eq. (\ref{eq:d}), parameters $q$ and $f$ as in Figs. \ref{fig:fig2} and \ref{fig:fig3}: (a) q-voter model -- $q=1$ (blue circles), $q=2$ (upward triangles), $q=3$ (diamonds), $q=4$ (squares), $q=5$ (downward triangles), $q=6$ (brown circles); (b) q-Ising model -- $q=1$ (blue circles), $q=2$ (upward triangles), $q=3$ (diamonds), $q=4$ (squares); (c) q-MV model -- $q=1$ (blue circles), $q=3$ (upward triangles), $q=5$ (diamonds), $q=4$ (squares) and (d) Sznajd model -- $f=1$ (blue circles), $f=0.75$ (upward triangles), $f=0.5$ (diamonds), $f=0.25$ (squares), $q=0.15$ (downward triangles). Lines are guidance to the eye.}
\label{fig:fig4}
\end{figure}

The results presented in Fig. \ref{fig:fig3} directly compare solutions of $F_r(m,a_1,r)=0$ with $F(m,a_1,a_2(r))=0$ for parameters $q$ (q-voter, q-Ising and q-MV) and $f$ (Sznajd) used previously in Fig. \ref{fig:fig2} and for specific values of $r$. The plots clearly show that at $r=0.1$, the solutions of $F$ and $F_r$ are almost indistinguishable, while at $r=0.5$, the similarity remains high. To describe the difference between the solutions of $F_r=0$ and $F=0$ in a more quantitative way, we use the relative distance (a relative $L^2$ error), defined as
\begin{equation}
    D(r) = \sqrt{\frac{\int_0^1\left[a_1(m)-a^{r}_1(m)\right]^2dm}{\int_0^1a^{r}_1(m)^2dm}}
    \label{eq:d}
\end{equation}
where $a_1(m)$ and $a_r(m)$ come from solving $F(m,a_1,a_2(r))=0$ $F_r(m,a_1,r)=0$ for $a_1$. Results of numerical integration presented in Fig. \ref{fig:fig4} confirm a high level of overlap between $F=0$ and $F_r=0$ solutions in Fig. \ref{fig:fig3}. In general, the values of $D$ are largely similar across all examined systems at specific values of $r$, grow monotonically with $r$, and models with larger $q$ are characterized by lower $D$. However, non-monotonic behavior is observed for q-Ising models other than $q=1$, whereas the q-voter at $q=6$ breaks the overall trend and yields the largest relative distance among all examined systems.

\section{Conclusions}
We introduced a general framework binding a two-layer system with heterogeneous dynamical parameters with a partially overlapped homogeneous one. The linking performed via the critical line has been shown to be sufficient to obtain a very good correspondence between these settings. We have tested this approach on several binary opinion models with diverse statistical properties (including continuous and discontinuous phase transitions) to demonstrate that the overall tendency is model-free. Surprisingly, for many systems, such as the q-Ising model, the q-voter with larger lobby sizes, or the Sznajd model, even relatively large values of overlap, such as $r=1/2$, still keep the relative distance at around 5\%. Although our results are derived in a strictly mean-field approach, i.e., for the full graph settings, some of the previous studies \cite{Jedrzejewski2017, Gradowski2020} indicate that for more structurally heterogeneous settings, the difference in behavior of the examined systems might not be substantial. 

\section*{Acknowledgments}
The authors of this work were funded by the European Union under the Horizon Europe grant OMINO - Overcoming Multilevel Information Overload (grant number 101086321, http://ominoproject.eu). Views and opinions expressed are those of the authors alone and do not necessarily reflect those of the European Union or the European Research Executive Agency. Neither the European Union nor the European Research Executive Agency can be held responsible for them. J.S. is financially supported by the Foundation for Polish Science (FNP) grant ‘Centre for Credible AI’ No. FENG.02.01-IP.05-0058/24
\bibliography{ref}

\end{document}